\documentclass[conference]{IEEEtran}
\IEEEoverridecommandlockouts
\usepackage{cite}
\usepackage{amsmath,amssymb,amsfonts}
\usepackage{algorithmic}
\usepackage{graphicx}
\usepackage{textcomp}

\usepackage[table,xcdraw]{xcolor}
\def\BibTeX{{\rm B\kern-.05em{\sc i\kern-.025em b}\kern-.08em
    T\kern-.1667em\lower.7ex\hbox{E}\kern-.125emX}}
\begin{document}

\title{Is a single unique Bayesian network enough to accurately represent your data?}

\author{\IEEEauthorblockN{1\textsuperscript{st} Gilles Kratzer}
\IEEEauthorblockA{\textit{Department of Mathematics, University of Zurich} \\
Zurich, Switzerland \\
gilles.kratzer@math.uzh.ch}
\and
\IEEEauthorblockN{2\textsuperscript{nd} Reinhard Furrer}
\IEEEauthorblockA{\textit{Department of Mathematics \&}
\IEEEauthorblockA{\textit{Department of Computational Science, University of Zurich}}
Zurich, Switzerland \\
reinhard.furrer@math.uzh.ch}
}

\maketitle

\begin{abstract}
  Bayesian network (BN) modelling is extensively used in systems epidemiology. Usually it consists
  in selecting and reporting the best-fitting structure conditional to the data. A major practical
  concern is avoiding overfitting, on account of its extreme flexibility and its modelling
  richness. Many approaches have been proposed to control for overfitting. Unfortunately, they
  essentially all rely on very crude decisions that result in too simplistic approaches for such
  complex systems. In practice, with limited data sampled from complex system, this approach seems
  too simplistic. An alternative would be to use the Monte Carlo Markov chain model choice
  ((MC)$^3$) over the network to learn the landscape of reasonably supported networks, and then to
  present all possible arcs with their MCMC support. This paper presents an R implementation, called
  \textsc{mcmcabn}, of a flexible structural (MC)$^3$ that is accessible to non-specialists.
\end{abstract}


\section{Introduction}

Bayesian Networks modelling is becoming more and more popular in systems epidemiology
\cite{b1,b2}. It is highly suitable for epidemiological datasets that are messy and highly
correlated or when it is not clear which variable would be the outcome. It is also well-adapted to
contexts where no prior model exists or when experts want a data-driven approach for selecting optimal
models. However, in highly interdisciplinary research fields, some concerns have been raised against
BN modelling.  A fully data-driven approach could select a model that is statistically supported by
the data but that does not have plausible epidemiological interpretation. To overcome this
limitation, there exists a popular workaround: one can either ban or retain some parts of the
structure to account for those biological constrains. This modelling approach thus ends up in a semi-supervised approach. Once this first guiding step is performed, all other arcs are either present or
absent in the model. 

From practice and modelling perspectives, a major concern of BN modelling is the tendency to
overfit the data and to select overly complicated models that generalise poorly. To compensate,
parametric or non-parametric approaches prune the selected structure. Usually, a unique and
theory-compatible structure is reported. Albeit being popular and accepted, this process makes very
crude choices regarding the possible connections in the model and diminishes the range of possible
interpretation. Indeed, an arc is either present or absent. This is exceptionally rudimentary considering the massive number of a priori networks.

%In contrast to other fields, the acceptance rate \rf{which?} of BN modelling, represented by directed acyclic
%graphs (DAGs), is relatively low compared to more traditional statistical approach within the
%epidemiological community. Even i
%In contexts where expert models are hard to identify or where more
%holistic approach would be preferable. Traditional regression based on expert model are preferred. A
%possible retraining factor is the high popularity of the p-values in epidemiology that act as
%proxies for significance and leads to rich and smooth interpretation of regression models. However
%being not recommended by many statistical associations, those tools are tentative to asses the
%plausibility of covariate in a model on a continuous scale. The p-values are certainly not the best
%answers but they are well accepted from an interpretation point of view. %\gk{note RF}

Classically in statistics, any relationship between variables is given with an estimate of the relationship based on data. A possible counterpart in BN modelling could be the Monte Carlo Markov chain
model choice ((MC)$^3$), which samples overall possible structures and moves from structure to
structure according to its support by the data. Indeed, the posterior probability of any
structural feature could be obtained from a Markov chain sample computed from graph structures by
\begin{equation}
E[f|D] \approx \frac{1}{S} \sum_{s=1}^{S}f(G^s), \text{ where } G^s \sim p(G|D)\label{eq}
\end{equation}
where $f$ is any structural query, $D$ is the data, $S$ is the set of visited structures $G$, and
$p(G|D)$ is the posterior distribution of the structures.

\section{Outline of the Approach}

Sampling BN using MCMC algorithms is a complicated task. The classical structural approach makes
single-edge operations (addition, deletion and reversal of an edge). It is known that naive
structural approaches will fail to efficiently sample BN landscapes in large problems. Indeed,
sampling algorithms mix slowly and hardly converge for large problem. A popular solution is to
sample from the space of node ordering using order MCMC sampling schemes \cite{b3}. But it is not
possible to explicitly express priors on graph structures in this context. This is of particular
importance in systems epidemiology where the data are usually scarce and, as a result, the posterior
depends heavily on the prior choice. Again, from a data analysis point of view, this would weaken
the analysis and render it not fully adapted. 

Two structurally transparent workarounds have been
proposed: the new edge reversal move \cite{b4} and Markov blanket resampling \cite{b5}. The former
advocates to make a reversal move in resampling the parent set. Indeed, the classical reversal move
depends on the global configuration of the parents and children, but then fails to propose MCMC
jumps that produce valid but very different DAGs in a unique move. The latter workaround applies the
same idea but to the entire Markov blanket of a randomly chosen node.  Single edge and both
structural methods were implemented in \textsc{mcmcabn}.

\section{Simulation Study}

To illustrate the usability of the approach for a typical systems epidemiology dataset, we simulate
a sparse BN of five nodes and five arcs (Fig.~\ref{fig:dag}).  250, 500, 1000, 10,000 binomial
observations were simulated from the DAG. A cache of scores for each possible set of parents was computed for each simulated dataset. The synthetic data was generated using the R package
\textsc{abn} \cite{b6}.  Afterwards, 10$^5$ sampling steps were performed for each
dataset. For computational reasons, the starting point of the MCMC chain is the true model and no
burn-in phase nor thinning has been considered. All computations were performed using R.

\begin{figure}[htbp]
  \centerline{\includegraphics[scale=0.53]{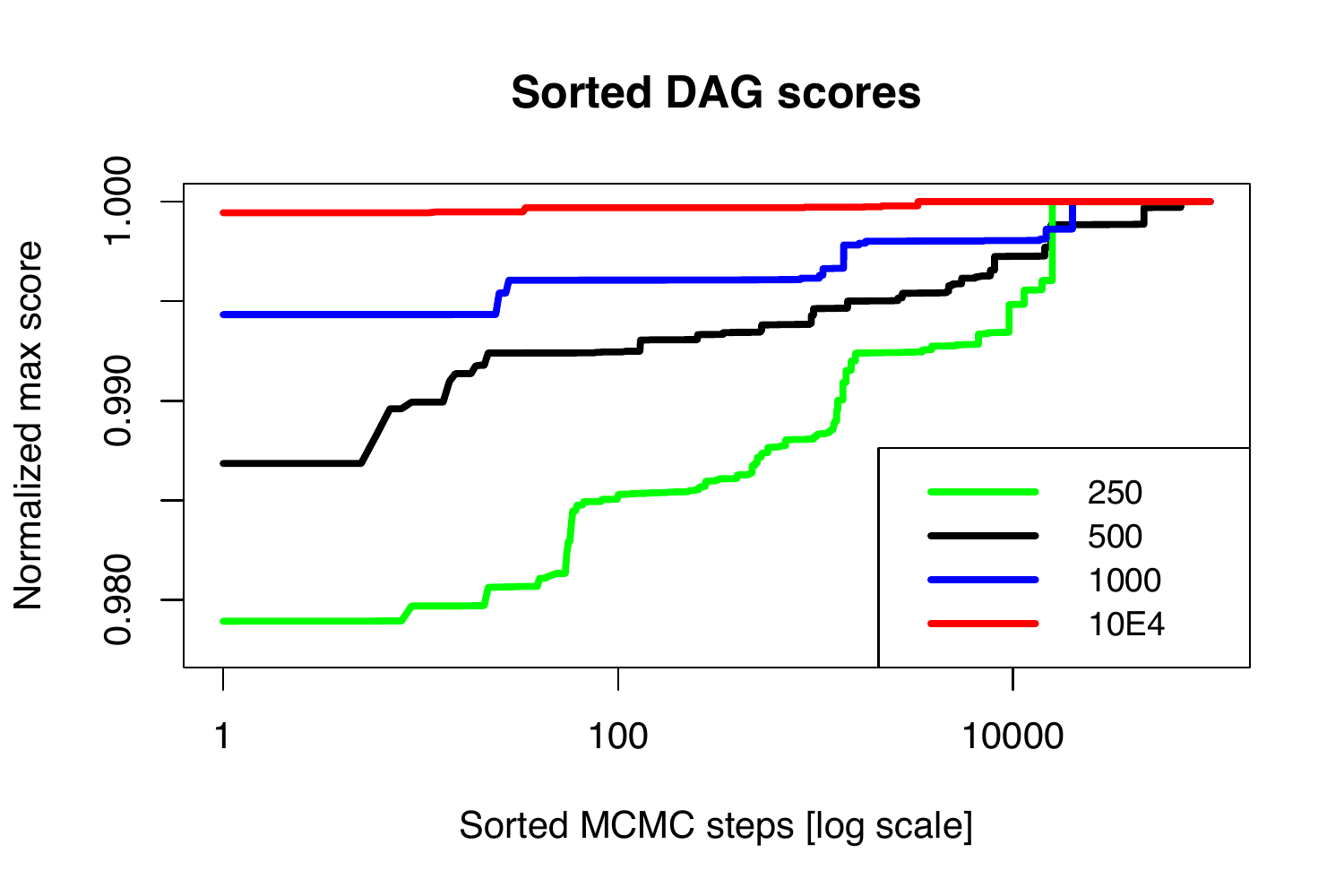}}
  \caption{Normalized maximal score in functions of the sorted simulated DAGs on a logarithmic scale
    for different sample sizes.}
  \label{fig1}
\end{figure}

Figure~\ref{fig1} illustrates the variability of the scores over the MCMC steps. For visual clarity,
the scores are normalized and the MCMC samples have been reordered (by increasing score). As dataset
size increases, less variability in term of structure is generated during the MCMC exploration. Because we
 started at the true model, we observe genuine sampling variability and not burn-in effects.

\begin{figure}[htbp]
  \centerline{\includegraphics[scale=0.53]{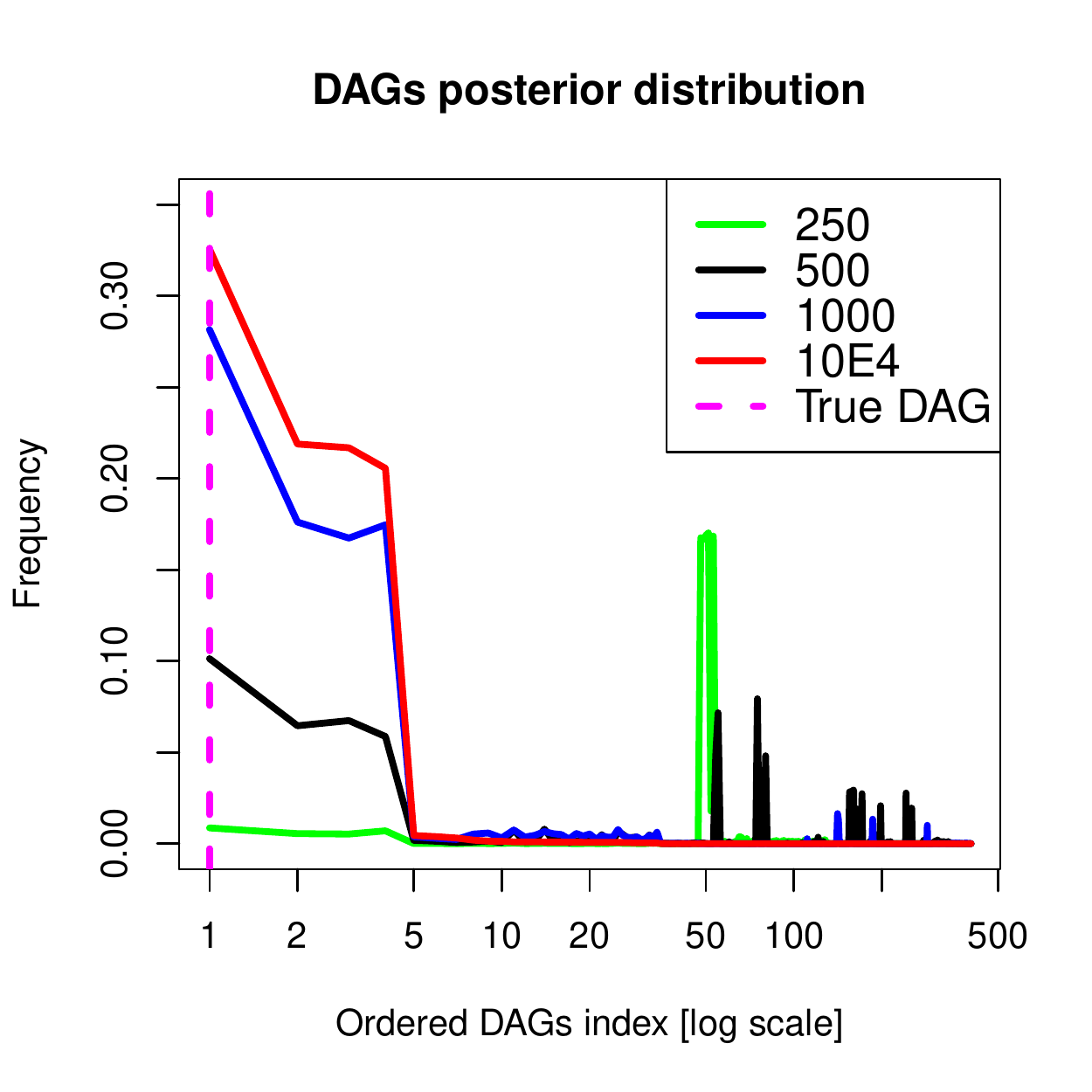}}
  \caption{Frequency of MCMC-generated DAGs for different sample sizes.}
  \label{fig2}
\end{figure}

Figure~\ref{fig2} shows the frequencies of sampled DAGs ordered by decreasing score of the largest
dataset. For this, the posterior consists almost exclusively of four DAGs of relatively similar
scores. The truth is captured in 1/3 of the cases.  However, 250 observations are not enough to
accurately sample the space of DAGs. 1000 observations lead to a substantial improvement of having
high frequencies of the dominant DAGs.

\section{Discussion}
\begin{figure}[h]
\begin{footnotesize}
\raisebox{-12mm}[2.5cm][1cm]{\includegraphics[scale=.38]{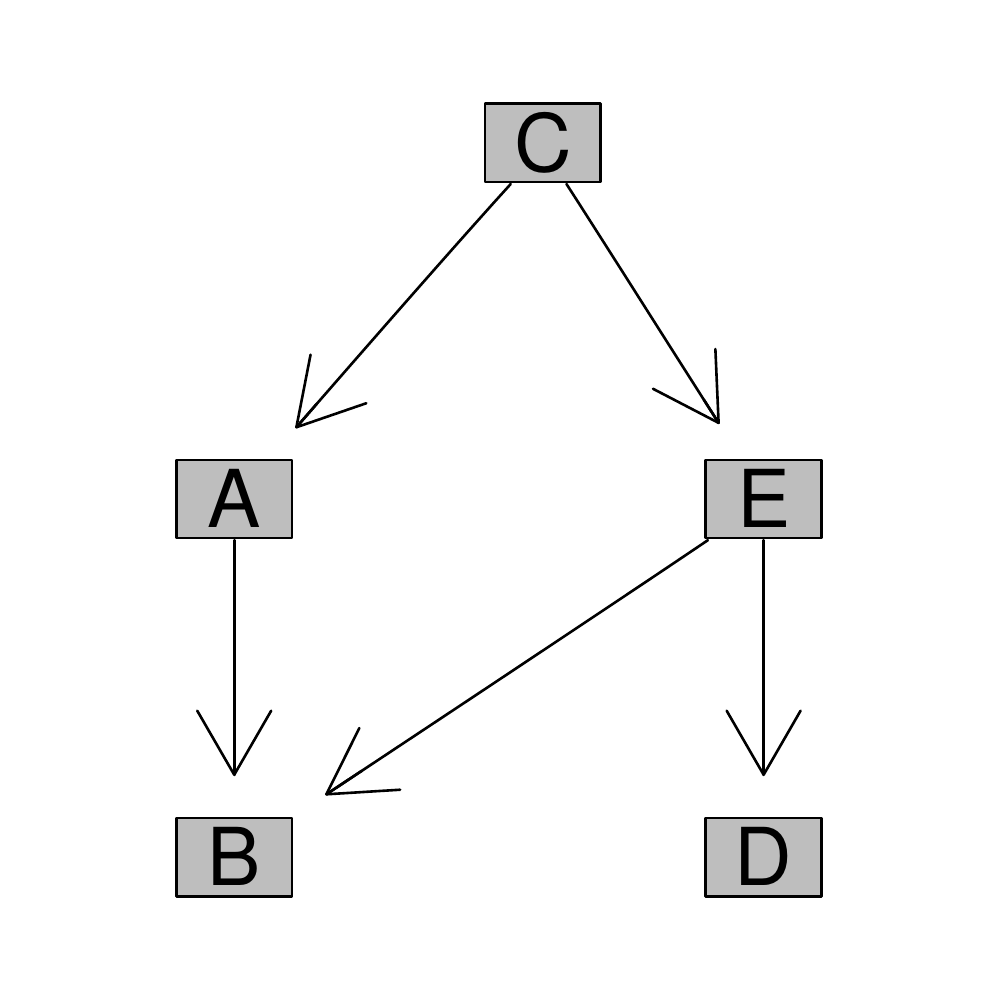}}
\tabcolsep4pt
\begin{tabular}{l|r|r|r|r}
\hline
\textbf{Queries}                & \multicolumn{1}{c|}{\cellcolor[HTML]{C0C0C0}\textbf{250}} & \multicolumn{1}{c|}{\cellcolor[HTML]{C0C0C0}\textbf{500}} & \multicolumn{1}{c|}{\cellcolor[HTML]{C0C0C0}\textbf{1000}} & \multicolumn{1}{c}{\cellcolor[HTML]{C0C0C0}\textbf{10000}} \\ \hline
\cellcolor[HTML]{FE0000}A - D & 1.8                                                     & 3.6                                                    & 1.3                                                      & 1.1                                                      \\ \hline
\cellcolor[HTML]{34FF34}A - B  & 5.3                                                     & 99.9                                                    & 100                                                        & 100                                                        \\ \hline
\cellcolor[HTML]{FE0000}B - C & 2.4                                                     & 34.0                                                    & 0                                                          & 0                                                          \\ \hline
\end{tabular}
\end{footnotesize}
\caption{Five-node DAG (left) and summaries of associated structural queries (\%) (right). Green indicates where the link is expected and red where it is not.} \label{fig:dag}
\end{figure}

The proposed approach makes it possible to find the optimal network by MCMC over DAG
structures. However, much more efficient algorithms have been proposed and are already implemented
in the R package \textsc{abn} \cite{b6}. Thus, the most desirable feature of the presented
approach is to enrich the analysis by computing structural queries over the DAG landscape.  Examples
of such queries are shown in Fig.~\ref{fig:dag}, where, not surprisingly, more observations lead to
better structural estimates. Here, the relationships between nodes are not directed.

\section*{Acknowledgments}
We would like to thank Fraser I. Lewis for the fruitful discussions about model averaging in BN modelling context that lead to this paper.

\vspace{12pt}


\begin{thebibliography}{00}
%\gk{first reference will be replaced by the second paper as soon as it will be accepted. I will say during the review round that "the first was replace by a more recent reference". Is it ok for you?} \rf{sure}
\bibitem{b1} S. Ruchti, A. R. Meier, H. W\"{u}rbel, G. Kratzer, S. G. Gebhardt-Henrich, S. Hartnack, ``Pododermatitis in group housed rabbit does in Switzerland--Prevalence, severity and risk factors'', Preventive Veterinary Medicine, vol. 158, pp. 114--121, October 2018.
\bibitem{b2} A. Comin, A. Jeremiasson, G. Kratzer, L. Keeling, ``Revealing the structure of the associations between housing system, facilities, management and welfare of commercial laying hens using Additive Bayesian Networks'', Preventive Veterinary Medicine, vol. 164, pp. 23-32, January 2019.
\bibitem{b3} N. Friedman, D. Koller, ``Being Bayesian about network structure'', Machine Learning, vol. 50, pp. 95--126, 2003.
\bibitem{b4} M. Grzegorczyk, D. Husmeier, ``Improving the structure MCMC sampler for Bayesian networks by introducing a new edge reversal move'', Machine Learning, vol. 71(2-3), pp. 265, 2008.
\bibitem{b5} C. Su, M. E. Borsuk, ``Improving structure MCMC for Bayesian networks through Markov blanket resampling'', The Journal of Machine Learning Research, vol. 17(1), pp. 4042-4061, 2016.
\bibitem{b6} G. Kratzer, M. Pittavino, F. I. Lewis and R. Furrer, ``abn: an R package for modelling multivariate data using additive Bayesian networks''. R package version 1.3.  https://CRAN.R-project.org/package=abn (2018)
\end{thebibliography}
\end{document}